\documentclass[epjST]{svjour}
\usepackage{graphicx}
\usepackage{color}
\usepackage{amssymb,amsmath,cite}

\begin{document}
\title{Creation and manipulation of entanglement in spin chains far from
equilibrium}
\author{F. Galve\inst{1,3},
        D. Zueco\inst{2,3}, 
        G.~M. Reuther\inst{3},
        S. Kohler\inst{3,4},
        and P. H\"anggi\inst{3}\fnmsep\thanks{\email{hanggi@physik.uni-augsburg.de}}}
\institute{IFISC (CSIC - UIB), Instituto de F\'{\i}sica Interdisciplinar y Sistemas 
  Complejos, Campus Universitat Illes Balears, 07122, Palma de
  Mallorca, Spain
\and
  Instituto de Ciencia de Materiales de Arag\'on y Departamento de F\'{\i}sica de la Materia Condensada, CSIC-Universidad de
  Zaragoza, 50009 Zaragoza, Spain
\and
  Institut f\"ur Physik, Universit\"at Augsburg,
  Universit\"atsstra{\ss}e~1, 86135 Augsburg, Germany
\and
  Instituto de Ciencia de Materiales de Madrid, CSIC, Cantoblanco,
  29049 Madrid, Spain}
%\institute{Insert the first address here \and the second here \and ...}
%
\abstract{
We investigate creation, manipulation, and steering of entanglement in 
spin chains from the viewpoint of quantum communication
between distant parties. We demonstrate how global parametric driving
of the spin-spin coupling and/or local time-dependent Zeeman fields
produce a large amount of entanglement between the first and the last
spin of the chain.  This occurs whenever the driving frequency meets a
resonance condition, identified as ``entanglement resonance''. Our
approach marks a promising step towards an efficient quantum state
transfer or teleportation in solid state system.  Following the
reasoning of Zueco \textit{et al.}~\cite{Zueco2009c}, we propose
generation and routing of multipartite entangled states by use of
symmetric tree-like structures of  spin chains. Furthermore, we study
the effect of decoherence on the resulting spin entanglement between
the corresponding terminal spins.}
\maketitle
\section{Introduction}
\label{intro}

Interacting spins are ubiquitous in physics, whether it be in magnetic
molecules~\cite{White1983a}, cold atoms~\cite{Bloch2008a}, Penning
traps~\cite{Ciaramicoli2008a} or Josephson-junction
arrays~\cite{Wendin2006a}. Since years, the equilibrium and
non-equilibrium properties of interacting spins have been extensively
studied. Phenomena such as quantum and classical phase transitions and
dynamical localization~\cite{Kenkre2000a} are well-known in the 
literature. The Heisenberg, Ising, and $XY$ coupling models became
paradigmatic models in condensed matter physics and statistical
physics~\cite{Nolting2009a}. 

Needless to say, the dynamics of interacting spins is of quantum
nature. The interest in those systems has been reinforced along with
new prospects in quantum technologies, which are expected to exceed
dramatically the capabilities of their classical analogs. Typical
examples are quantum metrology~\cite{Giovannetti2004a} or quantum
information processing~\cite{Amico2008}, with applications such as 
simulators or universal quantum computers. With the perspective of
quantum computing in mind, spin chains are natural connectors of
information between the different parts of the ``quantum hardware''
and the readout device~\cite{Bose2003a,Burgarth2006,Bose2006a}.

A key ingredient for all these applications is the entanglement
between  different spins in an array. Entanglement has no classical
analogue and marks the gap between classical and quantum technology.
Coming back to quantum communication protocols, the entanglement
shared by the emitter and receptor of the information ensures a
communication superior to what would be possible with classical
protocols~\cite{Bose2006a}. Consequently, the ability to fully control
entanglement creation and manipulation is a fundamental
requirement. One way to achieve this is 
by using time dependent fields. The benefits of external ac-driving
has been recently reported in Refs.~\cite{Zueco2009c, Galve2009a,
Burlak2009a, Leandro2009a, Wang2009a, diFranco2009a,
GarciaRipoll2007a}.  However, from a technological point of view, it
appears rather difficult to access each spin separately by applying
local magnetic fields.

In this work, we review and extend our recent results
\cite{Zueco2009c,Galve2009a} for entanglement control via global
external fields acting on all spins in an array.  Concretely, we
discuss entanglement creation between both ends of the chain, which
turns out to be optimal at particular driving resonances. Once
entanglement is created locally, neighboring spins interact such that
the spins at the ends of the chain efficiently communicate with each
other and entanglement routing through the chain remains the main
task. Here, we extend former studies on entanglement creation via
modulation of the spin-spin coupling~\cite{Galve2009a} to the case
where modulation of the local splittings is used instead. 

In the second part of this work, we report our idea of quantum
information routing by means of ac-fields. In detail, we propose a way
to produce multipartite entanglement between many distant parties just
by joining several ``quantum routers'' in ``beam-splitter''
mode. Finally, we provide studies on the validity of both driving
protocols discussed above under decoherence. 

The paper is organized as follows. In Sec.~\ref{sec:drivenspinchains}
we introduce the model and quantify the entanglement by means of the
concurrence between the outermost spins of the
chain. Section~\ref{sec:entanglementresonance} is devoted to
entanglement resonance, 
while in Sec.~\ref{sec:ac-control} we introduce the idea of routing
and entanglement distribution in quantum networks.  In
Sec.~\ref{sec:exp-decoh} we discuss a possible implementation in
Penning traps.

\section{Driven spin chains and entanglement computation}
\label{sec:drivenspinchains}

In this work we describe a chain of $N$ spins by the anisotropic
$XY$-model with time-dependent parameters using the
Hamiltonian~\cite{Lieb1961a} ($\hbar=1$)
 \begin{equation}
\label{Ham-gral}
H=\frac{1}{2}\sum_{n=1}^N h_n(t)
\sigma_n^z+\frac{J(t)}{4}\sum_{n=1}^{N-1}
\left[(1+\gamma)\sigma_n^x\sigma_{n+1}^x+(1-\gamma)\sigma_n^y\sigma_{n+1}^y\right]
\; .
\end{equation}
Both the local energy splittings $h_i(t)$ and the interaction
strengths $J(t)$ are assumed to depend through modulated global
fields acting on all spins at the same time. Nevertheless we allow for
inhomogeneities of the fields.  A non-zero value of the parameter
$\gamma$ refers to an homogeneous anisotropy.  For possible
applications in quantum communication the spin chain must have open
ends. Both the isotropic limit of the model $\gamma = 0$ and  the
thermodynamical limit $N \to \infty$ can be solved analytically in the
time-independent case by means of the Jordan-Wigner transformation
\cite{Lieb1961a,Mikeska1977}. On the contrary  we are interested in
both finite-size effects and time-dependent dynamics. We consider the
fields to be composed of dc contribution and one harmonic component,
such that
\begin{equation}\label{drive}
h_n(t)=\epsilon_n (h_0+h_1 \sin \omega t )\; , \quad
J(t)=J_0+ J_1 \sin \omega t .
\end{equation}
For later convenience we move to an interaction picture defined by
$\widetilde X = U_0^\dagger(t) X U_0(t)$, where $U_0(t) = \exp[-i
\sum_n \varphi_n (t)\sigma^z_n]$ and $\dot \varphi_n(t) = h_n(t)/2$.
Introducing the ladder operators $\sigma^\pm = \frac{1}{2} ( \sigma^x
\pm {\rm i} \sigma^y)$, the Hamiltonian~\eqref{Ham-gral} gets replaced
by the interaction-picture Hamiltonian
%-------------------------
\begin{equation}
\begin{split}
\label{Ham-gral-int}
\widetilde H=
\frac{1}{2} \big(J_{0} + J_1 \, \sin(\omega t) \big)
\sum_{n=1}^{N-1}
\Big[ &
\sigma_n^+\sigma_{n+1}^- {\rm e}^{{\rm i}  \Delta_n (t)}
+
\sigma_n^-\sigma_{n+1}^+ {\rm e}^{-{\rm i}  \Delta_n (t)}
\\ &
+
\gamma \Big (
{\rm e}^{{\rm i}  \Sigma_n (t)}
\sigma_n^+\sigma_{n+1}^+ +
{\rm e}^{-{\rm i}  \Sigma_n (t)}
\sigma_n^-\sigma_{n+1}^-
\Big)
\Big] ,
\end{split}
\end{equation}
%----------------------------
where $\Delta_n = \varphi_n - \varphi_{n+1}$ and $\Sigma_n =
\varphi_n + \varphi_{n+1}$.
The first two terms in the sum of Hamiltonian~\eqref{Ham-gral-int}
swap the states of spins $n$ and $n+1$, whereas the latter two terms
create and destroy excitations, respectively. As it will turn out, the
swapping terms are responsible for entanglement transfer while the
latter pair of terms constitutes the origin of entanglement
generation.

In the following, we are interested in the bipartite entanglement
between two spins in the chain. For this purpose one needs to obtain
the reduced density matrix of spins $j$ and $k$. It results from the
full density matrix of the chain $\varrho_{\rm tot}$ by tracing out
all other degrees of freedom,
\begin{equation}
\varrho_{jk}(t) = {\rm Tr}_{n\neq j,k} 
    \big [ \varrho_\mathrm{tot} (t) \big ] \;,
    \quad n \neq j,k \; ,
\end{equation}
where  ${\rm Tr}_{n\neq j,k}$ denotes the partial trace with respect to all
spins $n=1,2,\ldots N$ but spins $j$ and $k$. The entanglement shared by
spins $j$ and $k$ is then given by the
concurrence~\cite{Wootters1998a}
\begin{equation}\label{concurrence}
C_{jk}=\max\{\lambda_1-\lambda_2-\lambda_3-\lambda_4,0\} \, .
\end{equation}
The $\lambda_\nu$ are the ordered square roots of
the eigenvalues of the matrix $\varrho_{jk}(\sigma_j^y \sigma_k^y)
\varrho_{jk}^* (\sigma_j^y \sigma_k^y)$.

\section{Resonances and entanglement creation}
\label{sec:entanglementresonance}

The time dependence of the coefficients in the
Hamiltonian~\eqref{Ham-gral} generally leads to a complex dynamics, in
addition to the nontrivial dynamics for the time-independent case.
Therefore, we confine ourselves to scenarios in which both amplitudes
$J_1$ and $h_1$ from Eq.~\eqref{drive} are smaller than the dc field
strength $h_0$. We focus on the spectral response of the chain to
periodically modulated coupling. All other parameters have arbitrary
but fixed values. For simplicity, we assume in this section a
homogeneous chain, $\epsilon_n = 1$ for all $n$. 

In the protocol, the spins are initially uncoupled, i.e.\ $J(t)=0$ for
$t<0$, and cooled down to the fully-aligned separable state
\begin{equation}
\label{psi0}
| \psi(t{=}0)\rangle  = |0000\ldots \rangle \, ,
\end{equation}
which is the ground state of the Hamiltonian~\eqref{Ham-gral} with
$J=0$. At $t=0$, we switch on both the coupling $J(t)$ and the
sinusoidal driving $h(t)$. This procedure is termed a quantum quench
and constitutes a typical protocol for entanglement
generation~\cite{Wichterich2009a}. Even though the interaction
strength is usually considered as constant, we take into account the
possibility of time-dependent coefficients $J_1(t) >0$ for $t>0$. In
this sense we are introducing time-periodic quantum quench dynamics.

In the following, we raise the question of how much entanglement
between the ends, measured in terms of the concurrence, may be created
of the chain. Such large-distance entanglement could be used for
typical quantum communication protocols, such as teleportation or
state transfer~\cite{Bose2006a}.

\subsection{Time-dependent interaction}
\label{sec:timedep-interact}

At first, we only consider the interaction to depend on time,
i.e.\ $h_1 = 0$ and $J_1 \neq 0$. We investigate the time evolution of
the concurrence 
between spin zero ($n=0$) and the last spin ($n=N$) for different
driving frequencies $\omega_d$, and the parameters $\gamma$, $J_1$ and
$J_0$ by direct numerical integration. Concerning the impact of the
chain length $N$ we refer to Ref.~\cite{Galve2009a}.
We determine the maximal
concurrence $C_{1,N}^{\mathrm{max}}$ that can be obtained within the
time interval $T=[0,\ldots,4 N/\max(J_0,J_1)]$. The results depicted
in Fig.~\ref{fig:resonant-dr}(a) reveal that at $\omega_d = 2 h_0$ the
maximal concurrence assumes a value close to unity during this time
interval and is significantly larger than for other frequencies. We
refer to this resonance condition as
\textit{entanglement resonance}, which holds independently of the
other parameters. On the contrary, we find that the height and width
of the resonance peak depend on the intensities $J_0$ and $J_1$ and
on the chain length $N$ (not shown in the figure, see
Refs.~\cite{Galve2009a}). We also notice the 
existence of a secondary, smaller peak at the driving frequency
$\omega_d=h_0$.  In contrast to the main peak, its amplitude strongly
decreases with the coupling intensity and increasing chain
length~\cite{Galve2009a}.

In order to understand this phenomenon quantitatively, we
replace the Hamiltonian~\eqref{Ham-gral-int} by its time average
within a rotating-wave approximation (RWA). At the resonance condition
$\omega_d = 2 h_0$, we obtain
\begin{equation}
\widetilde {H}_{R} =
\frac{J_{0}}{2}\sum_{n=1}^{N-1} \Big[
   \sigma_n^+\sigma_{n+1}^-
 + \tilde \gamma \sigma_n^+\sigma_{n+1}^+ + \text{h.c.}
\Big] ,
\label{effective}
\end{equation}
with the effective anisotropy $\tilde \gamma =\gamma J_1/2J_0$. This
means that for resonant driving, the time-dependent $XY$-model
\eqref{Ham-gral} can be mapped to the static
$XY$-model~\eqref{effective} without Zeeman fields. In both cases,
the entanglement generated between the two ending spins is maximal and
controlled by the parameter $\widetilde \gamma$ and the chain length
$N$. Note that $J_0\to 0$ corresponds to the
infinitely anisotropic
limit $\tilde\gamma\to\infty$. One important point is worth being
mentioned:  the concurrence approaches unity in the limit of vanishing
$J_0$, i.e.\ for infinite $\tilde \gamma$. In this limit, the amount of
entanglement no longer depends on $J_1$ and $\gamma$. An argumentation given
in Ref.~\cite{Galve2009a} explains why the limit $J_0 \to 0$ is
optimal for entanglement generation. As a rule of thumb, the
concurrence becomes maximal when minimizing the swap terms in the
Hamiltonian~\eqref{Ham-gral-int}.
Let us finally mention that in Ref.~\cite{Galve2009a}, the
entanglement scalability with the chain length $N$ and the arrival
time for this entanglement was studied.

%%%%%%%%%%%%%%%%%%%%%%%%%%%%%%%%%%%%%%%%%%%%%%%%%%%%%%%%%%%%%%%%%%%%%%%
%%%%%%%%%%%%%%%%%%%%%%%%%%%%%%%%%%%%%%%%%%%%%%%%%%%%%%%%%%%%%%%%%%%%%%%

\subsection{Modulation of the local fields}
\label{sec:local-fields}

\begin{figure}
\resizebox{0.98\columnwidth}{!}{
  \includegraphics{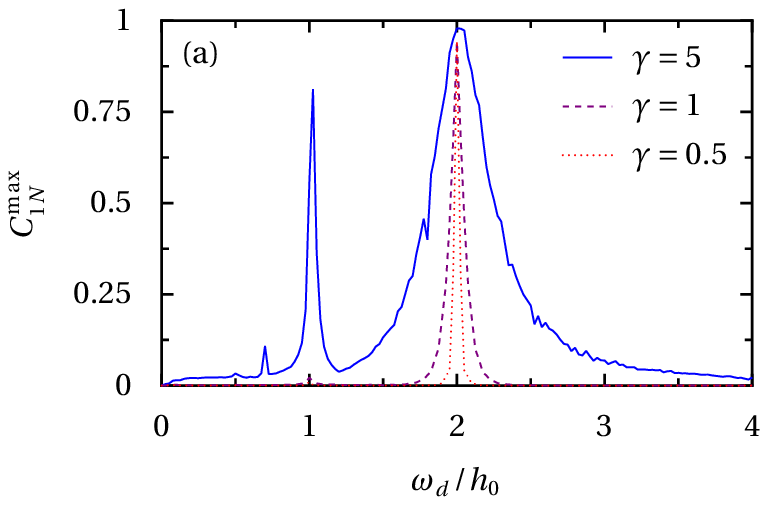}\hspace{1ex}
  \includegraphics{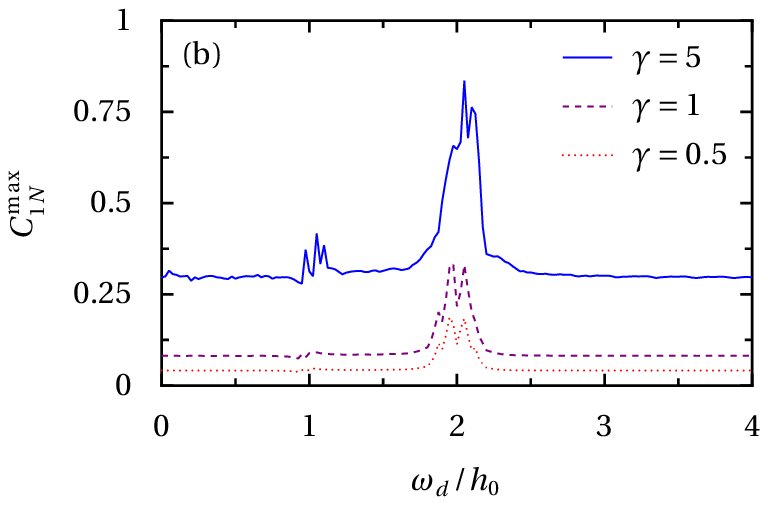}}
\caption{Maximal concurrence  $C_{1N}^{\mathrm{max}}$ between the end 
  spins as a function of the driving frequency for various
  anisotropies. (a) Entanglement creation by modulation of the
  interaction $J(t)$ with the parameters $h_1=0$, $J_0=0$, and 
  $J_1=0.1 h_0$, cf.\ Eq.~\eqref{drive}. (b) Corresponding effect due to
  modulation of the Zeeman fields $h_n (t)$, for $h_1=0.1 h_0$,
  $J_0=0.1 h_0$, and $J_1=0$. The chain length is $N=6$ in both
  cases.}  
\label{fig:resonant-dr}
\end{figure}

In the following, we investigate the response to driving with
local fields. We assume that at $t=0$ the interaction is switched
to a constant value with $J_0 >0, J_1 = 0$.  However, the 
local Zeeman field has now both ac and dc contributions, $h(t) = h_0 +
h_1 \sin (\omega t)$. Again we assume a homogeneous chain, $\epsilon_j
= 1$ for all $j$. Proceeding exactly as above, we compute the maximum
concurrence reachable within the time interval $[0,\ldots,4
N/\max(J_0,J_1)]$. Our numerical results are drawn in
Fig.~\ref{fig:resonant-dr}(b). Also in this case we find an
entanglement resonance for $\omega_d = h_0$ and for $\omega_d = 2
h_0$, where the peak at $\omega_d = 2 h_0$ dominates clearly. In order
to understand this behavior we recall the Hamiltonian in the
interaction picture, Eq.~\eqref{Ham-gral-int}, simplified by setting
$J_1 = 0$ in this case,
%-------------------------
\begin{equation}
\label{He-gralpm-int-RWA}
H=
\frac{J_{0}}{2}
\sum_{n=0}^{N-1}
\Big [
\sigma_n^+\sigma_{n+1}^- + \sigma_n^-\sigma_{n+1}^+
+
\gamma \Big (
{\rm e}^{{\rm i}  \Sigma (t)}
\sigma_n^+\sigma_{n+1}^+ +
{\rm e}^{-{\rm i}   \Sigma (t) }
\sigma_n^-\sigma_{n+1}^-
\Big )
\Big]
\end{equation}
%----------------------------
with $\Sigma(t) = 2 \int_0^t {\rm d}s ( h_0 + h_1 \sin(\omega_d t))$.
Notice that the driving no longer affects the swapping terms.  On the
other hand, the entanglement in terms of the concurrence is favored
by the terms $\sigma_n^+ \sigma_{n+1}^+  + {\rm h.c}$, as pointed out
after Eq.~\eqref{Ham-gral-int}. For this reason, the resonances should
assume values for which the time-dependent coefficients
$\exp [{\rm i}  \Sigma (t)]$ are not averaged to zero. To get
further insight, we perform a Taylor expansion in powers of ($h_1 /
\omega$), yielding
\begin{equation}
{\rm e}^{{\rm i}  \Sigma (t)}=
{\rm e}^{{\rm i} h_1/\omega_d}
\sum_{n}
(-{\rm i})^{n}
\left (
\frac{h_1}{\omega_d}
\right)^{n}
\frac{1}{n!}
\sum_{k}
\left (
\begin{array}{c}
n
\\
k
\end{array}
\right )
{\rm e} ^{ {\rm i} [2h_0 + (n - 2k) \omega_d] t}
\label{effective-h}
\end{equation}
where $\tbinom nk = n!/k!(n-k)!$ are binomial coefficients. With this
at hand we obtain the resonance condition
%------------------
\begin{equation}\label{rescond-bin}
2h_0 + (n - 2k) \omega_d=0
\end{equation}
%------------------
To lowest order, Eq.~\eqref{rescond-bin} is fulfilled for $n=1$,
which yields $\omega_d = 2 h_0$. This again confirms
our numerical results depicted in Fig.~\ref{fig:resonant-dr}. The next
order at $n=2$ yields the second resonance condition $\omega_d = h_0$,
which explains the second peak in the figure. Peaks corresponding to
resonances of higher orders ($n>2$) are suppressed for $h_1/\omega_d
\ll 1$. 

Finally, we emphasize that the
swapping  terms $\sigma_n^+ \sigma_{n+1}^-$ are unaffected by driving
the local fields. Thus, less concurrence is obtained as compared to the
case with time-dependent interaction described in
Sec.~\ref{sec:timedep-interact}. This feature becomes apparent when
comparing the Hamiltonians~\eqref{effective} and~\eqref{effective-h}
or, equivalently, Figs.~\ref{fig:resonant-dr}(a) and
\ref{fig:resonant-dr}(b).

%%%%%%%%%%%%%%%%%%%%%%%%%%%%%%%%%%%%%%%%%%%%%%%%%%%%%%%%%%%%%%%
%%%%%%%%%%%%%%%%%%%%%%%%%%%%%%%%%%%%%%%%%%%%%%%%%%%%%%%%%%%%%%%

\section{Directing of entanglement current via ac-control}
\label{sec:ac-control}

\begin{figure}
 \centerline{\includegraphics[scale=.99]{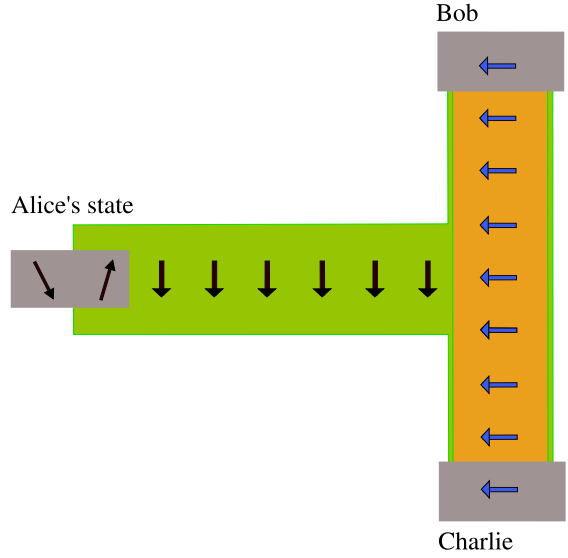}}
\caption{Model for the simplest quantum router possible.  A green spin
  chain allows perfect transfer and is not driven, while the
  propagation in the orange chain is controlled by ac fields.
}
\label{fig:router}
\end{figure}

\begin{figure}
  \centerline{\includegraphics[scale=.99]{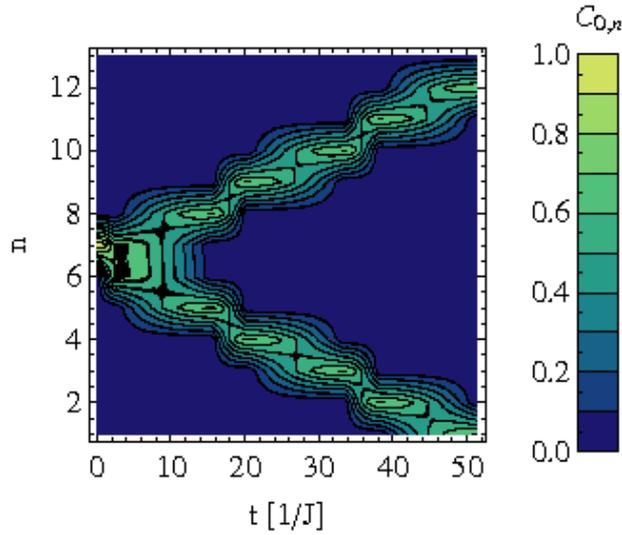}}
\caption{Entanglement transport in a spin chain structure
  corresponding to Fig.~\ref{fig:router}. Concurrence $C_{0,n}$
  [Eq.~\eqref{concurrence}] between the spin at Alice's site
  in the horizontal chain  (labeled
  here $n=0$) and the spins in the vertical chain. The
  entanglement is split up, arriving both Bob ($n=12$) and Charlie
  ($n=1$) after a given time. Here, the chain is 
  isotropic ($\gamma=0$), while the Zeeman fields
  have a ratchet-like globally driven shape as described in
  Ref.~\cite{Zueco2009c}.}   
\label{fig:concurrencesplit}
\end{figure}

In Ref.~\cite{Zueco2009c} the intriguing possibility of entanglement
steering in an isotropic $XY$ chain ($\gamma=0$) was explored.  It was
shown how to make excitations in the spin chain propagate into a
specific direction with a ratchet-like profile of the local Zeeman
fields~\cite{Hanggi2009a} plus a superimposed oscillation. This
suggests the fabrication of a ``quantum router''.  In this way,
distant parties could be brought to share a high amount of
entanglement, provided that decoherence is sufficiently small.  As a
further option, the excitation may be split into two parts, each
propagating to one end of the chain. This results in entangling three
distant parties rather than two. In Fig.~\ref{fig:router} we
sketch a model for the simplest router possible, where the
entanglement has two possibilities of propagation. Corresponding
numerical results about splitting of entanglement into two directions
are depicted in Fig.~\ref{fig:concurrencesplit}.

To extend and generalize this elementary sketch of a quantum router,
we point out how multipartite entanglement could be created
between many distant parties. For this purpose, we suggest choosing a
setup in which the entanglement signal has been sent through a
driven spin chain, split in two. Both tail ends of the chain get thus
entangled with the sender of the packet. The initial state
${|}100\rangle$ having been sent by Alice, the final state becomes
\begin{equation}\label{AlBobCh}
| \text{Alice, Bob, Charlie} \rangle =
\frac{1}{\sqrt{2}}\Big[|100\rangle
+\frac{1}{\sqrt{2}}\big(|010\rangle+|001\rangle\big)
\Big]
\; ,
\end{equation}
and is now shared by Alice, Bob and Charlie (the entries of the state
vectors are given in this order). State~\eqref{AlBobCh} is a
tripartite entangled state with similarity to a
$W$-state~\cite{Dur2000a,Walther2005a}, except for the weighting
factors. It represents the different quantum coherent histories, i.e.\
outcome possibilities of excitation transfer.  The component
${|}100\rangle$ refers to the case in which Alice has not sent any
excitation through the driven chain. The latter two terms
${|}010\rangle$ and ${|}001\rangle$ result from splitting an
excitation, which has been sent by Alice before.

\begin{figure}
\centerline{\includegraphics[scale=0.99]{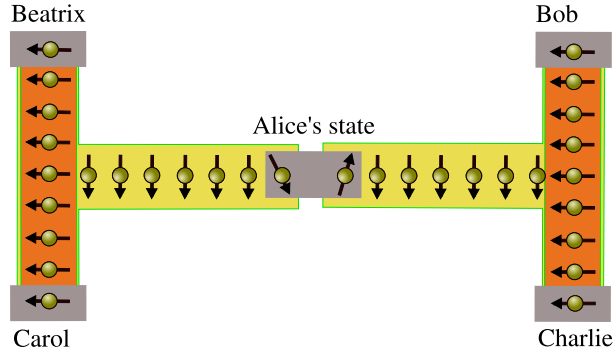}}
\caption{Setup for the production of quadripartite
  distant entangled state, where Alice sends here initial singlet into
  two different driven chains which both are in operated in the splitting mode.
}
\label{fig:multipart}
\end{figure}

\begin{figure}
\centerline{\includegraphics[scale=0.99]{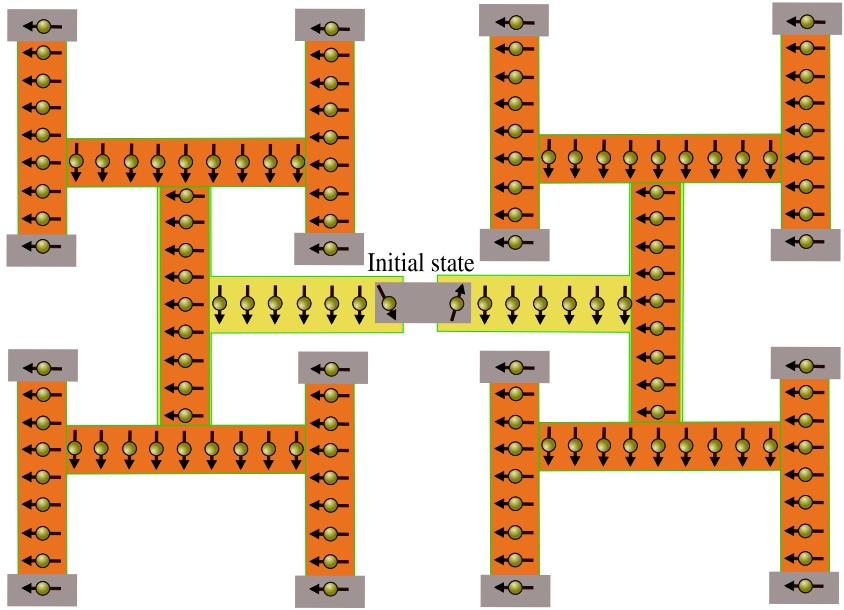}}
\caption{Setup for the production of multipartite (sixteen parties)
 distant entangled states where Alice's singlet is split several
 times. As above the gray boxes mark output terminals.
}
\label{fig:multipart-big}
\end{figure}

If other driven chains are appended to Bob and Charlie, initial
excitations can be split further. This results in a final
penta-partite entangled state, i.e.\ a state held between five parties,
which exhibits yet two more terms. However, the share of the
distant parties gets diluted by a factor $1/\sqrt{2}$ every time the
signal is split, while Alice's fraction never gets diminished. To
get over this drawback, Alice can send her excitation to another
spin array in a different direction. As illustrated in
Fig.~\ref{fig:multipart}, this would result in a quadripartite state,
where Bob and Charlie are entangled with Beatrix and Carol,
\begin{equation}
| \text{Beatrix, Carol, Bob, Charlie} \rangle =
\frac{1}{2}\big[|1000\rangle+|0100\rangle+|0010\rangle+|0001\rangle\big]
\end{equation}
This state is well balanced, and further splittings keep this balance
unaffected if performed symmetrically on both ends. We notice too
that the cardinality of the states is now even, whereas  it is odd in
the above scenario. To exploit advantages of multipartite distant
entangled states, one could use a setup such as sketched in
Fig.~\ref{fig:multipart-big}. 

\section{Experimental realizations and decoherence effects}
\label{sec:exp-decoh}

A highly accurate and stable confinement of ions in traps has been
achieved in the last decade, together with the possibility of
operating in the ground state of the spatial ionic
motion~\cite{Roos1999a,wineland2003a}. Several proposals to use this
amazing tools in order to simulate spin chains have been
given~\cite{Ciaramicoli2008a,Porras2003}. Here, the use of magnetic
field gradients and state dependent laser forces, respectively, yield
an effective spin-spin coupling mediated by the spatial degrees
of freedom. The latter is achieved through Coulomb repulsion between
ions~\cite{Ciaramicoli2003}, or a wire-mediated capacitive coupling of
the ions residing in neighboring traps~\cite{Stahl2005a}.
These proposals allow for the accurate simulation of many spin
Hamiltonians, included Eq.~\eqref{Ham-gral}, provided the  
mean photon numbers (temperatures) with respect to the relevant 
degrees of freedom are low. It must be noted that
in~\cite{Ciaramicoli2008a} spins as such are used indeed, whereas
in~\cite{Porras2003} they are simulated by means of two internal
electronic ion levels. The latter proposal has been experimentally
realised~\cite{Porras2008}. Furthermore, the first two Fock levels of
the axial motion can be used to implement spins for the purpose of
quantum routing~\cite{Zueco2009c}, given that the ions in the chain
can be prepared in the ground state with a probability of
$99.9\%$~\cite{Roos1999a}.  

In the following, we take into account decoherence, which is the main
obstacle for the creation of entanglement through resonant driving. It
originates from weak but unavoidable coupling of the spin chain to a
typically large number of uncontrollable external degrees of
freedom. Irrespective of its physical nature, the environment is
usually modeled as a bath of harmonic oscillators, each coupling to a
system coordinate via its position
operator~\cite{Magalinskii1959a,Caldeira1983a,Grifoni1998a,Hanggi2005b}. 
Here we assume that each qubit $n$ undergoes pure and independent
dephasing, i.~e. that its coordinate $\sigma_n^z$ couples to a
separate bath.

For weak dissipation, one can eliminate the bath within second-order
perturbation theory and derive a Bloch-Redfield master equation for
the reduced density operator $\rho$ of the qubits~\cite{Blum1996a}.
Considering only phase noise on the traps, the equation takes a
Lindblad form
\cite{Lindblad1976a},
\begin{equation}
\label{qme}
\dot{\rho}
= -\frac{{\rm i}}{\hbar}[H,\rho]
 +\frac{\lambda}{2} \sum_{n=1}^N (\sigma_n^z\rho\sigma_n^z-\rho),
\end{equation}
where we have assumed that the effective decoherence rate $\lambda$ is
the same for all qubits.  The above
master equation has been obtained perturbatively in the dissipation
strength.  However, it turns out to be valid beyond the perturbative
regime and therefore provides an accurate description of the dephasing 
dynamics~\cite{Doll2008a}. The heating rate, which would include
relaxation terms in the master 
equation, is typically several orders of magnitude smaller than the
dephasing rate. This justifies the modelling of decoherence by pure
phase noise. We use a rather conservative estimate and assume
$\lambda\sim10^{-4}J$. 

\begin{figure}
\resizebox{0.98\columnwidth}{!}{
 \includegraphics{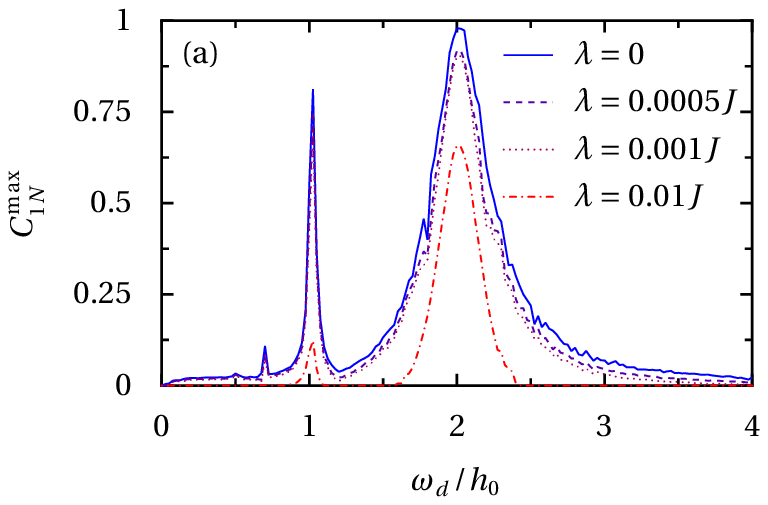}\hspace{1ex}
  \includegraphics{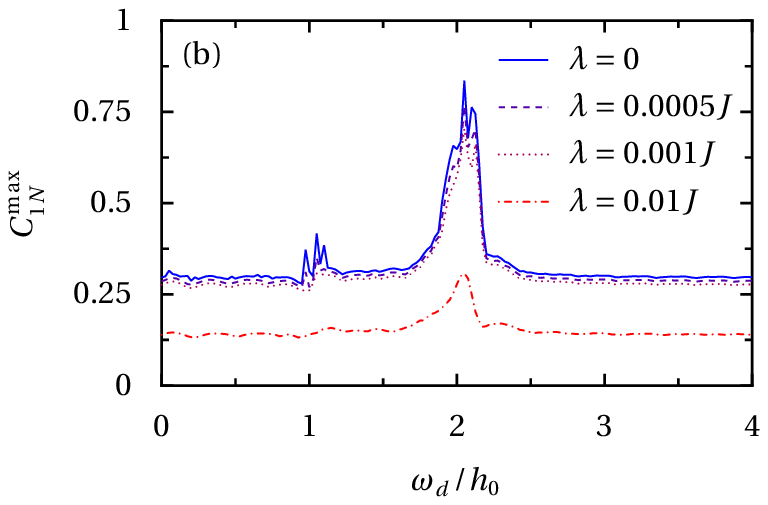} }
\caption{Maximal concurrence $C_{1N}^{\rm max}$ for resonant driving
  [Eq.~\eqref{drive}] as a function of the driving frequency for
  various decoherence rates.  The anisotropy parameter is $\gamma=5$,
  while the chain length is $N=6$. (a)
  Modulation of the interaction $J (t)$ with the parameters
  $h_1=0$, $J_0=0$, and $J_1=0.1 h_0$. (b)
  Modulation of the fields $h_n (t)$, where $h_1=0.1 h_0$, $J_0=0.1
  h_0$, and $J_1=0$.} 
\label{fig:decoherence}
\end{figure}

Now we consider the impact of decoherence in the resonance phenomena 
discussed in Sec.~\ref{sec:entanglementresonance}. In
Fig.~\ref{fig:decoherence} we plot the concurrence versus the
frequency at different decoherence rates. Obviously, the concurrence
becomes smaller with increasing decoherence.  This reduction is
related to the quantum master equation~\eqref{qme} with respect to the
time it takes for the entanglement signal to reach the ends of the
chain.  Even so,  the entanglement is well preserved for a decoherence
rate $\lambda = 0.001 J$. This is quite a realistic estimate for most 
physical systems that are suitable for experimental verification of
entanglement resonance. For this reason, it should be possible to
observe entanglement transport in an experiment as proposed, even
though it will be limited by decoherence. 

\section{Conclusions}
\label{sec:conclusions}

The transfer of methods established in condensed matter physics to the
field of quantum information, and vice versa, has recently undergone a
blossoming and promoting activity~\cite{Schenk2007a}. With this work,
we have reviewed 
and extended our proposals \cite{Zueco2009c, Galve2009a} for
generating and routing entanglement in a spin chain by utilizing 
time-dependent control of spin-spin interactions. Entanglement routing is
achieved by a global manipulation of the chain via an external driving
fields. This method possesses a salient advantage as compared to the
rather difficult local addressing of individual spins, since that task
requires a delicate tuning of intrinsic system parameters. An optimal
amount of entanglement is generated at a particular resonance, which
builds a bridge to the familiar theory of  coherent destruction of
tunneling~\cite{Grifoni1998a,Grossmann1991a,Grossmann1991b}. 

Thus, driving with external fields paves the way to control
entanglement dynamics in realistic spin chains.  We further elaborated
on the role of decoherence thereby incorporating the unavoidable
influence of environmental degrees of freedom in a realistic scenario.
With regard to an experimental realization, we have sketched the
implementation with Penning traps. Another possibility consists in
using tailored Josephson junction arrays.

A further advantage of our time-dependent driving scheme is the
substantial decoherence suppression by use of
ac-fields~\cite{Grifoni1998a,Grifoni1995a,Grifoni1996a,Kurizki2002a}.
This time-dependent manipulation allows the optimal control in
attaining maximal entangled of distant objects~\cite{Galve2009a}. In
addition, Landau-Zener dynamics has been proposed as well for
entanglement generation protocols~\cite{Wubs2007a,Zueco2010a}.
Finally, the implementation of optimal quantum gates was
suggested~\cite{FonsecaRomero2005a}. All in all, this justifies the
efforts in employing the toolboxes from both driven quantum systems
and solid state correlation physics in administrating quantum
information processing.

\medskip
{\small
The authors thank the Deutsche Forschungsgemeinschaft for financial
support via the collaborative research center SFB
484. F.G. acknowledges support by COQUSYS. D.Z. acknowledges partial
support from FIS2008-01240 and FIS2009-13364-C02-01 (MICINN).
}

\bibliographystyle{epj}

%\bibliography{../gr_bib/decoherence}

\begin{thebibliography}{45}

\bibitem{Zueco2009c}
D.~Zueco, F.~Galve, S.~Kohler, P.~H{\"a}nggi, Phys. Rev. A \textbf{80}, 042303
  (2009)

\bibitem{White1983a}
R.M. White, \emph{Quantum theory of magnetism}, 2nd~edn. (Springer, Berlin,
  1983)

\bibitem{Bloch2008a}
I.~Bloch, J.~Dalibard, W.~Zwerger, Rev. Mod. Phys. \textbf{80}, 885 (2008)

\bibitem{Ciaramicoli2008a}
G.~Ciaramicoli, I.~Marzoli, P.~Tombesi, Phys. Rev. A \textbf{78}(1), 012338
  (2008)

\bibitem{Wendin2006a}
G.~Wendin, V.S. Shumeiko, in \emph{Handbook of Theoretical and Computational
  Nanotechnology}, edited by M.~Rieth, W.~Schommers (American Scientific
  Publishers, Los Angeles, 2006), Vol.~3, p. 223

\bibitem{Kenkre2000a}
V.M. Kenkre, J. Phys. Chem. B \textbf{104}, 3960 (2000)

\bibitem{Nolting2009a}
W.~Nolting, \emph{Quantum Theory of Magnetism} (Springer, New York, 2009)

\bibitem{Giovannetti2004a}
V.~Giovannetti, S.~Lloyd, L.~Maccone, Science \textbf{306}, 1330 (2004)

\bibitem{Amico2008}
L.~Amico, R.~Fazio, A.~Osterloh, V.~Vedral, Rev. Mod. Phys. \textbf{80}, 517
  (2008)

\bibitem{Bose2003a}
S.~Bose, Phys. Rev. Lett. \textbf{91}, 207901 (2003)

\bibitem{Burgarth2006}
D.~Burgarth, Ph.D. thesis, University College London (2006), arXiv:0704.1309
  [quant-ph]

\bibitem{Bose2006a}
S.~Bose, Contemp. Phys. \textbf{48}, 13 (2007)

\bibitem{Galve2009a}
F.~Galve, D.~Zueco, S.~Kohler, E.~Lutz, P.~H\"anggi, Phys. Rev. A \textbf{79},
  032332 (2009)

\bibitem{Burlak2009a}
G.~Burlak, I.~Sainz, A.B. Klimov, Phys. Rev. A \textbf{80}, 024301 (2009)

\bibitem{Leandro2009a}
J.F. Leandro, A.S.M. de~Castro, F.L. Semi{\~a}o, ArXiv:0909.3545 [quant-ph]
  (2009)

\bibitem{Wang2009a}
X.~Wang, A.~Bayat, S.G. Schirmer, S.~Bose, ArXiv:0911.5405 [quant-ph] (2009)

\bibitem{diFranco2009a}
C.~DiFranco, M.~Paternostro, M.S. Kim, ArXiv:0911.5160 [quant-ph] (2009)

\bibitem{GarciaRipoll2007a}
O.~Romero-Isart, J.J. Garc{\'{\i}}a-Ripoll, Phys. Rev. A \textbf{76}, 052304
  (2007)

\bibitem{Lieb1961a}
E.~Lieb, T.~Schultz, D.~Mattis, Ann. Phys. \textbf{16}, 407 (1961)

\bibitem{Mikeska1977}
H.J. Mikeska, W.~Pesch, Z. Phys. B \textbf{26}, 351 (1977)

\bibitem{Wootters1998a}
W.K. Wootters, Phys. Rev. Lett. \textbf{80}, 2245 (1998)

\bibitem{Wichterich2009a}
H.~Wichterich, S.~Bose, Phys. Rev. A \textbf{79}, 060302(R) (2009)

\bibitem{Hanggi2009a}
P.~H\"anggi, F.~Marchesoni, Rev. Mod. Phys. \textbf{81}, 387 (2009)

\bibitem{Dur2000a}
W.~D{\"u}r, G.~Vidal, J.I. Cirac, Phys. Rev. A \textbf{62}, 062314 (2000)

\bibitem{Walther2005a}
P.~Walther, K.J. Resch, A.~Zeilinger, Phys. Rev. Lett. \textbf{94}, 240501
  (2005)

\bibitem{Roos1999a}
C.~Roos, T.~Zeiger, H.~Rohde, H.C. N{\"a}gerl, J.~Eschner, D.~Leibfried,
  F.~Schmidt-Kaler, R.~Blatt, Phys. Rev. Lett. \textbf{83}, 4713 (1999)

\bibitem{wineland2003a}
D.~Leibfried, R.~Blatt, C.~Monroe, D.~Wineland, Rev. Mod. Phys. \textbf{75},
  281 (2003)

\bibitem{Ciaramicoli2003}
G.~Ciaramicoli, I.~Marzoli, P.~Tombesi, Phys. Rev. Lett. \textbf{91}, 017901
  (2003)

\bibitem{Stahl2005a}
S.~Stahl, F.~Galve, J.~Alonso, S.~Djekic, W.~Quint, T.~Valenzuela, J.~Verdu,
  M.~Vogel, G.~Werth, Eur. Phys. J. D \textbf{32}, 139 (2005)

\bibitem{Magalinskii1959a}
V.B. Magalinskii, Zh. Eksp. Teor. Fiz. \textbf{36}, 1942 (1959), [Sov. Phys.
  JETP {\bf 9}, 1381 (1959)]

\bibitem{Caldeira1983a}
A.O. Caldeira, A.L. Leggett, Ann. Phys. (N.Y.) \textbf{149}, 374 (1983)

\bibitem{Grifoni1998a}
M.~Grifoni, P.~H\"anggi, Phys. Rep. \textbf{304}, 229 (1998)

\bibitem{Hanggi2005b}
P.~H\"anggi, G.L. Ingold, Chaos \textbf{15}, 026105 (2005)

\bibitem{Blum1996a}
K.~Blum, \emph{Density Matrix Theory and Applications}, 2nd~edn. (Springer, New
  York, 1996)

\bibitem{Lindblad1976a}
G.~Lindblad, Commun. Math. Phys. \textbf{48}, 119 (1976)

\bibitem{Doll2008a}
R.~Doll, D.~Zueco, M.~Wubs, S.~Kohler, P.~H{\"a}nggi, Chem. Phys. \textbf{347},
  243 (2008)

\bibitem{Porras2003}
D. Porras and J. I. Cirac, Phys. Rev. Lett. \textbf{92}, 207901 (2004)

\bibitem{Porras2008}
A. Friedenauer, H. Schmitz, J. T. Glueckert, D. Porras and T. Schaetz,
Nat. Phys. \textbf{4}, 757 (2008)

\bibitem{Schenk2007a}
S.~Schenk, G.L. Ingold, Phys. Rev. A \textbf{75}, 022328 (2007)

\bibitem{Grossmann1991a}
F.~Grossmann, T.~Dittrich, P.~Jung, P.~H\"anggi, Phys. Rev. Lett. \textbf{67},
  516 (1991)

\bibitem{Grossmann1991b}
F.~Grossmann, P.~Jung, T.~Dittrich, P.~H\"anggi, Z. Phys. B \textbf{84}, 315
  (1991)
  
\bibitem{Grifoni1995a}
M.~Grifoni, M.~Sassetti, P.~H\"anggi, U.~Weiss, Phys. Rev. E \textbf{52}, 3596
  (1995)

\bibitem{Grifoni1996a}
M.~Grifoni, M.~Sassetti, U.~Weiss, Phys. Rev. E \textbf{53}, R2033 (1996)

\bibitem{Kurizki2002a}
G.~Kurizki, A.G. Kofman, D.~Petrosyan, T.~Opatrn{\'y}, Journal of Optics B
  \textbf{4}, 294 (2002)

\bibitem{Wubs2007a}
M.~Wubs, S.~Kohler, P.~H\"anggi, Physica E \textbf{40}, 187 (2007)

\bibitem{Zueco2010a}
D.~Zueco, G.M. Reuther, P.~H{\"a}nggi, S.~Kohler, Physica E \textbf{42}, 363
  (2010)

\bibitem{FonsecaRomero2005a}
K.M. Fonseca-Romero, S.~Kohler, P.~H\"anggi, Phys. Rev. Lett. \textbf{95},
  140502 (2005)

\end{thebibliography}
%
\end{document}